\documentclass[%
superscriptaddress,
twocolumn,
10pt,
amsmath,amssymb,
aps,
pra,
floatfix,
]{revtex4-1}

\usepackage{graphicx}
\usepackage{color,soul} 
\usepackage{dcolumn}
\usepackage{bm}
\usepackage{multirow}
\usepackage{nicefrac}

\newcolumntype{d}[1]{D{.}{.}{#1}}

\usepackage{hyperref}
\begin{document}
\title{Practitioner's guide to social network analysis: Examining physics anxiety in an active-learning setting}

\author{Remy Dou}
\email[]{redou@fiu.edu}
\affiliation{Department of Teaching and Learning, Florida International University, Miami, Florida 33199}
\affiliation{STEM Transformation Institute, Florida International University, Miami, Florida 33199}
\author{Justyna P. Zwolak}
\email[]{j.p.zwolak@gmail.com}
\affiliation{Joint Center for Quantum Information and Computer Science, UMD, College Park, MD 20742, USA}

\date{\today}
 
\begin{abstract} 
The application of social network analysis (SNA) has recently grown prevalent in science, technology, engineering, and mathematics education research. Research on classroom networks has led to greater understandings of student persistence in physics majors, changes in their career-related beliefs (e.g., physics interest), and their academic success. In this paper, we aim to provide a practitioner's guide to carrying out research using SNA, including how to develop data collection instruments, setup protocols for gathering data, as well as identify network methodologies relevant to a wide range of research questions beyond what one might find in a typical primer. We illustrate these techniques using student anxiety data from active-learning physics classrooms. We explore the relationship between students' physics anxiety and the social networks they participate in throughout the course of a semester. We find that students' with greater numbers of outgoing interactions are more likely to experience negative anxiety shifts even while we control for {\it pre} anxiety, gender, and final course grade. We also explore the evolution of student networks and find that the second half of the semester is a critical period for participating in interactions associated with decreased physics anxiety. Our study further supports the benefits of dynamic group formation strategies that give students an opportunity to interact with as many peers as possible throughout a semester. To complement our guide to SNA in education research, we also provide a set of tools for letting other researchers use this approach in their work -- the {\it SNA toolbox} -- that can be accessed on GitHub. 
\end{abstract}

\maketitle

\section{Introduction}\label{sec:intro}
The principle that information exists within, and because of, human interactions with one another anchors many theories of philosophy, sociology, and knowledge development~\cite{Bourdieu86-FC,Lin99-SC,Borgatti09-NSS,NRC2000}. Even the knowledge that exists within our scientific enterprises, however objectively we approach our research questions, has to go through a series of socially constructed hurdles before finding acceptance in our communities. The peer-review process exemplifies that. For that reason, social scientists, including education researchers, have began to study the nature of interactions between people and how those interactions facilitate (or hinder) information flow and development. 

The way social interactions affect learning experiences can vary significantly between individuals. For example, some students like discussing their ideas to reaffirm their knowledge. They may face little difficulty when reaching out to others for help or to offer support. As such, they truly thrive in an environment that promotes peer-to-peer and student-faculty interactions. Others dread sharing their ideas in public, especially when these ideas are still developing. It may be because of a sense of anxiety, a feeling of self-consciousness, or shyness. Whatever the reason, such students might have difficulties appreciating active-engagement learning strategies and even get discouraged from persisting in a course. Understanding how and why students build communities, and how these communities affect their educational well-being is essential to improving their experiences in and beyond the classroom.

One way to approach this problem is to examine student integration using the tools of social network analysis (SNA). While SNA does not directly capture the content of interactions, it allows us to quantify the various aspects of relational structures that result from those interactions~\cite{Scott2011}. The application of SNA has recently grown prevalent in science, technology, engineering, and mathematics (STEM) education research. From classroom network dynamics and career persistence to school-level group belonging and information sharing, network methodology has proven itself useful in helping researchers understand factors affecting students’ success in STEM~\cite{Cho08-CIS,Zwolak17-IIP,Zwolak18-PIN}. However, while there are resources for those interested in the application of SNA, the few primers that exist fail to provide enough detail to carry out nuanced education studies and the more in-depth textbooks lack a classroom framework by which to interpret results from such analyses~\cite{Grunspan14-SNA,Scott2011,Bruun13-GNF,Carolan2014}. A succinct, higher-level practical guide showing the entire process from designing relevant tools to collecting data to applying SNA in educational contexts is (to the best of the authors' knowledge) currently absent from literature. This work is intended to fill in this gap.

For over half a decade we have applied SNA in the field of physics education research. That work has led to greater understandings of student persistence, changes in their career-related beliefs (e.g., physics interest), and their academic success~\cite{Zwolak17-IIP,Zwolak18-PIN,Dou16-BPM,Dou18-USE,Williams17-EPP}. In the process, we also established SNA study design in the classroom context, including development of data collection instruments, setup of protocols for gathering and digitizing data, as well as identification of network methodologies relevant to a wide range of research questions. We also built a software suite -- the {\it SNA toolbox} -- that allows to carry the network analysis presented in the following sections. In this paper, we aim to present these approaches and techniques in SNA using the context of student anxiety, and to discuss how outcomes and interpretations vary based on methodological and analytical choices. We focus on social networks found in classrooms, i.e., networks representing peer-to-peer and student-instructor interactions. Our goal is to provide a succinct guide that remains practical to the education researcher exploring classroom, departmental, or institution-related interactions between people, regardless of the specific question being examined. 

This is not intended to be a primer. Rather, this paper will delve into the nuanced aspects of social network analysis, providing guidance along the way that goes beyond a basic explanation of a few centrality measures, and will address considerations when collecting data, performing analyses, and interpreting outcomes. Finally, we focus solely on the context of the physics classroom, using our research of student anxiety in an active-learning setting to illustrate the content. Nevertheless, the applications of the SNA topics addressed here, as well as the provided {\it SNA toolbox} code~\cite{sna-tools}, can be used in other physics education research contexts (and beyond).

The paper is organized as follows: After a brief overview of research on anxiety in the introductory physics classroom (Sec.~\ref{sec:framework}) and after introducing the physics anxiety survey (Sec.~\ref{sub:anx-survey}), we proceed to the first major section: the ``Social network analysis survey'' (Sec.~\ref{sub:sna-survey}). This section addresses questions one should consider when determining data collection context, survey development, administration of surveys, and handling of multiple collections. In particular, we discuss what constitutes social interactions and how one can measure them. We then introduce different types of social networks and present guidance on developing surveys that yield the network type of interest. We also introduce measures that can be used to examine weighted network data, as well as guidelines for their interpretation within the classroom context (e.g., what does it mean for a student to have high ``closeness'' centrality, Sec.~\ref{sub:sna-tools}). Finally, we also discuss practical aspects of data collection: the administration of surveys, handling of multiple collections, accounting for non-normality of data and handling missing data (Sec.~\ref{sub:other}). The statistical analysis techniques that we use are presented in Sec.~\ref{sub:stats}. The second major section, ``Analysis and results'' (Sec.~\ref{sec:analysis}), shows practical applications of the proposed methodologies in the context of students' physics anxiety in active-learning introductory physics courses. We conclude with a discussion of our findings, limitations of this work, and recommendations for future directions in Sec.~\ref{sec:summary}.

To make the discussed methodologies more user-friendly, we established a GitHub repository where we make available the R source-code together with a manual and a simple reproducible example that can be easily adapted and used to carry out SNA analyses (open source, available at GitHub~\cite{sna-tools}). While presently the {\it SNA toolbox} includes only code used in the analysis from this manuscript, it will be continuously maintained and extended further based on the needs of and requests from the science education community.

\section{Anxiety in the introductory physics classroom}\label{sec:framework}
To explore the relationship between physics anxiety and in-class student interactions in an active-learning setting, we adopt a participationist framework. Participationists primarily view learning as ``the development of ways in which an individual participates in well-established communal activities''~\cite{Sfard01-MTE}. Learning is perceived as a construction of mutual understandings within a social context, with emphasis placed on examining discourse and interactions rather than ``acquisition'' of knowledge as a commodity or object~\cite{Sfar98-TML}. As such, we espouse the philosophy that ``learning and social interactions are not mutually exclusive''~\cite{Dou16-BPM}.

Our motivation to focus on physics anxiety is predicated on our belief that anxiety shapes how and to what extent students participate in classroom activities. If physics anxiety hinders participation, our framework suggests learning, too, may suffer. Prior work in the realm of social anxiety -- not physics anxiety, per se -- has found negative correlations with participation in activities that may be present in active-learning settings. For example, it has been suggested that social anxiety leads to risk averse behaviors as individuals seek to preserve how others perceive their image or identity~\cite{Schlenker92-SAM}. Such behaviors can lead to reticence or complete unwillingness to present before an audience, particularly in settings framed around the evaluation of content being shared. Active-learning curricula often nurture these kinds of settings, where students publicly present results to one another. Even when public presentations are not directly related to evaluation, the perception of being evaluated can have an impact on behavior~\cite{Paivio59-MCA}. Hills calls out constructivist teaching styles in particular~\cite{Hills07-ICR}. In his study of pre-service math and science teachers, he found that those with high social anxiety tended to exhibit risk aversion behavior, which manifested in the classroom as low group participation and avoidance of open-ended math problems. Even the productivity of group-brainstorming has been shown to be negatively affected by the level of groups members' social anxiety~\cite{Camacho95-AGB}.

The correlation between various types of anxiety and physics learning at the undergraduate level have been documented by several researchers. Williams found that students who reported feeling anxious about communicating in class, even in non-group, whole-class settings (e.g., when an instructor poses a question to the class) were less likely to score well on multiple-choice exams and less likely to exhibit large gains on the Force Concept Inventory (FCI)~\cite{Hestenes92-FCI,Williams01-UAG}. Engineering students' math anxiety while learning electricity and magnetism has been shown to be negatively correlated with course exam scores, as well as conceptual understanding~\cite{Leppavirta11-PHD}. 

The idea that, in addition to communication and math anxiety, physics anxiety should be considered as a unique construct that affects physics learning is over thirty years old and has been associated with studies related to gender differences in physics learning~\cite{Tobias85-TA}. More recently, Sahin~\cite{Sahin14-PTA} explored the physics anxiety of pre-service teachers pursuing careers in science, math, and primary education (e.g., physics education, secondary math education) who were at the time enrolled in an introductory physics course. Outcomes of this study showed that those in the physics education program exhibited less anxiety than those in any of the other programs, but found that significant gender differences existed for physics-focused majors, such that female pre-service physics teachers were more likely to exhibit higher physics anxiety than their male counterparts. The study also found that students with high physics anxiety tended to have earned either low (i.e., $<2$) or high (i.e., $>3$) GPA, which the author admits runs contrary to related literature that essentially admonishes for a linear, indirect relationship between anxiety and performance. 

The relationship between anxiety, participation, and student outcomes motivated our exploration of the potential social mechanisms through which it manifests in an active-learning, student-centered classroom. As described earlier, past research identifies participation in academic activities as a factor of student anxiety. Thus, we expect students' physics anxiety to have a negative effect on their participation. We also take into account past research identifying social support as a mitigator of anxiety~\cite{Crockett07-SSA}. We thus expect to find a relationship between changes in anxiety and students' social embeddedness within the classroom network, such that students who seek out relationships with their peers will be more likely to feel less anxious about physics over time. We also hypothesize that the frequency with which students carry out repeated interactions with the same individuals exhibits a weaker relationship with anxiety than the number of unique individuals a student interacts with (i.e., having a greater number of people to provide possible support). Additionally, our analyses take into account students' self-reported gender and final course grade.

\section{Methodology}\label{sec:methods}
In this section, we present the Physics Anxiety Rating Scale (PARS)~\cite{Sahin15-PARS} and the social network survey we use to collect data for this study. We also discuss some of the considerations we took into account when designing our examination of physics anxiety through a social network lens. For completeness, we include the ``Social Network Analysis Toolbox'' section that presents network measures we rely on when comparing data between different groups and sections. While not exhaustive, this list is intended to give flavor for what kind of information can be extracted and quantified using SNA. 

\subsection{Physics anxiety survey}\label{sub:anx-survey}
To measure students' physics anxiety, we use a 16-item version of the PARS developed by Sahin~\cite{Sahin15-PARS}. The PARS asks students to rate their agreement with a variety of statements on a 5-item Likert scale ranging from strongly disagree (1) to strongly agree (5). The statements include the following: ``I would feel very embarrassed if the instructor corrected the answer that I gave to a physics question in front of the class'', ``being unable to use units of quantities appropriately in physics courses makes me very anxious'', and ``when solving a physics problem, I worry about not being able to recall relevant formulas or physics laws''. The survey data is typically collected in {\it pre} and {\it post} format, using the same instrument at the beginning and end of the semester, and allows to capture changes in anxiety. The Cronbach's alpha reliability coefficient is 0.92 on the scale using {\it pre} data and 0.94 using {\it post} data. 

Since all students coming to the class are expected to have some level of anxiety~\cite{Tilgner90-ASE}, which typically varies across individuals, we are interested in the anxiety shift rather than the raw anxiety score. As the semester goes on and students experience the curriculum, we expect to see an increase or decrease in their anxiety score, depending on their learning experiences. We avoid ascribing value to initial student anxiety (e.g., high anxiety is bad, low anxiety is good) since such practices can conflict with past research indicating that certain levels of anxiety positively correlate with quality performance~\cite{Eysenck07-ACP}. 

To provide a measure of the anxiety shift over time, we calculate the normalized anxiety shift defined as the ratio of the absolute shift to the maximum possible shift~\cite{Hake98-MIP}:

\begin{equation}\label{eq:nor-shift}
s_{norm} = \frac{post - pre}{max.\,\,possible.\,\,score - pre},
\end{equation}
where {\it pre} and {\it post} denote the score of a student on the anxiety survey before and after the course, respectively. This approach allows a comparison of shifts between students with varying {\it pre} scores. The {\it max. possible. score} on this survey is $80$ and the lowest possible score is $16$. Note that, as this measure was developed to assess the expected average gains on the FCI (i.e., positive shifts averaged over the entire class), it is not robust against dramatic drops in scores of individuals. In particular, for the PARS survey the $s_{norm}$ will be outside of the $[-1,1]$ boundaries when the {\it pre} score is over 48 and the {\it post} score is lower than $2(pre-40)$ (see Appendix~\ref{app:nor_gain} for more details). As such, the $s_{norm}$ can be used to identify potential ``outliers'' -- a medium to high pre-course score followed by an unexpectedly low {\it post} anxiety may identify students who did not offer reliable responses on the {\it post}-survey. After careful considerations it might be advisable to either remove the unusually low {\it post} scores and impute the missing data or to remove such individuals from analyses all together.

\subsection{Social network analysis survey}\label{sub:sna-survey}
Identifying a relevant theoretical framework prior to designing social network research provides boundaries and guidance for the measurement instrument (e.g., survey design), analysis (e.g., correlational study), and interpretation. Here we discuss the design of the SNA survey that we use to gauge classroom participation. Like any other research tool, SNA should be applied only when the context of a study makes it an appropriate tool. In what follows, we discuss when SNA is the right method of analysis, what constitutes social interactions, and how one can measure them. We also discuss the practical aspects of data collection, including administration of surveys (e.g., on-line vs. in-person, one-time vs. longitudinal collection) together with brief analyses of pros and cons for each approach. The main purpose of this section is to present guidance on developing surveys that yield accurate and relevant networks. To illustrate the process we explain how our study meets the requirements. During the design of the SNA survey and its administration, we carefully considered our responses to all questions posted below.

\subsubsection*{\texorpdfstring{$\mathcal{Q}1$}{Q1}: \bf\emph{Is SNA an appropriate tool to help answer my question?}}

To use SNA, the research question(s) have to be related to interactions of some kind, be it students working in groups, e-mail exchanges, participation in a forum, or co-authoring a paper, to name a few. In our study, we want to explore how engagement in a student-centered physics classroom contributes to anxiety shifts while also taking pre-course anxiety into account. Our focus on student-student interactions lends itself to quantification via SNA. 

\subsubsection*{\texorpdfstring{$\mathcal{Q}2$}{Q2}: \bf\emph{What interactions am I interested in?}}

\begin{figure}[t]
\centering
\includegraphics[width=0.9\linewidth]{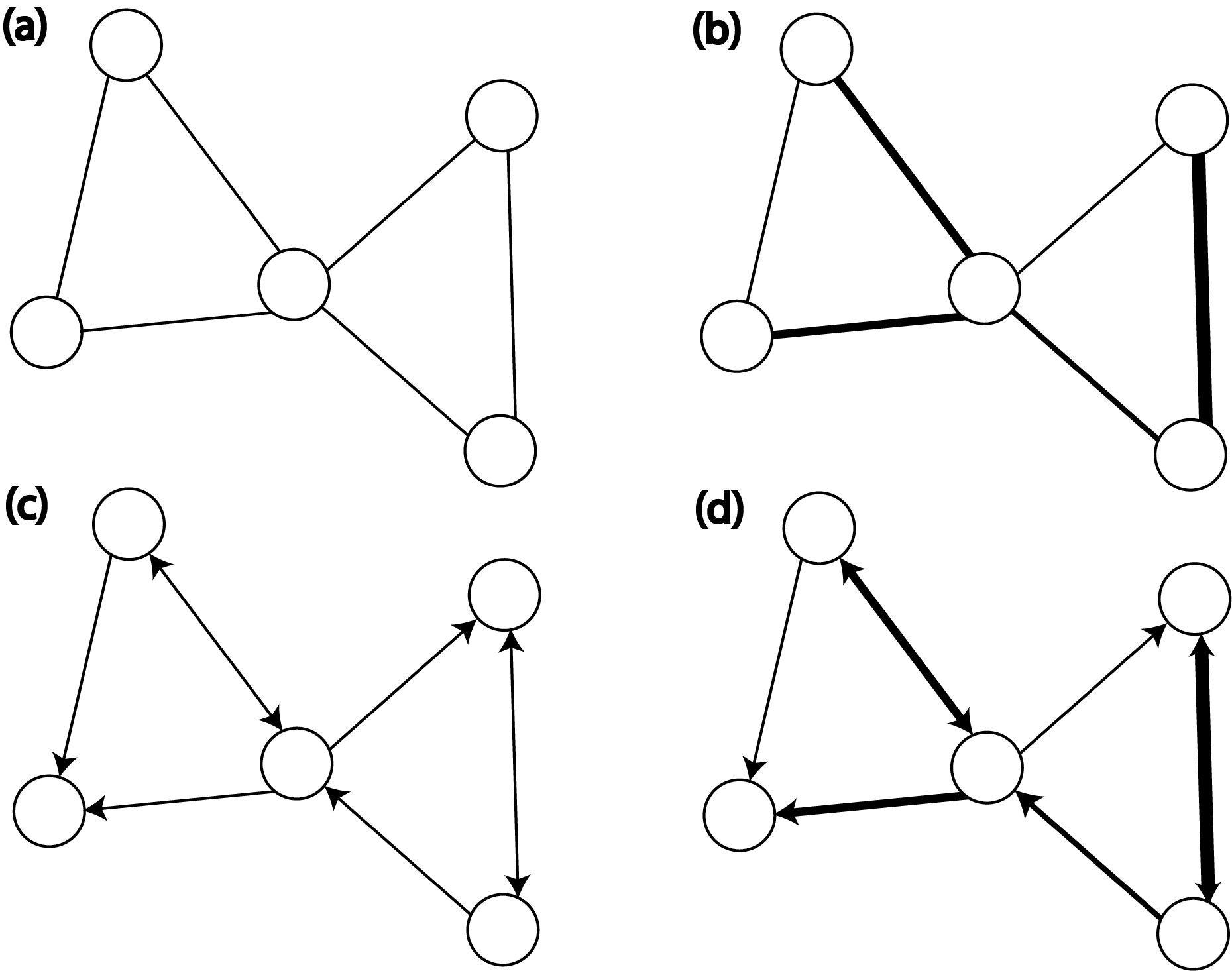}
\caption{Visualization of various types of networks: (a) binary and undirected, (b) weighted and undirected, (c) binary and directed, (d) weighted and directed.\label{fig:nets}}
\end{figure}

Although the context of a study helps to establish how interactions should be defined (e.g., conversations, joint papers, participating in the same meetings, etc.), one needs to decide early on what additional characteristics of interest to incorporate. For instance, is it important to know who initiated a given interaction (i.e., directed vs.~undirected networks, see Fig.~\ref{fig:nets}(a) and (c))? Is it important to know how frequently a given interaction occurred or does it only matter whether it took place (i.e., weighted vs.~binary networks, see Fig.~\ref{fig:nets}(a) and (b))? Whose perspective matters -- the initiator's or the receiver's? Should all members of a particular group be included in the network? 

For our study, we define ``interaction'' as a meaningful (from a respondent's perspective) in-class interaction related to physics. This may include, among other behaviors, a discussion of ideas, joint work on a problem, as well as listening to others solve or discuss problems. We also want to know the frequency of interactions between the same two individuals in a given week.  Thus, we opt to collect directed network data that captures the frequency with which the interactions take place within a given collection period (see Fig.~\ref{fig:nets}(d) for a visualization of this type of network). This grants us flexibility during the analyses to calculate centrality measures that place more or less emphasis on both directionality and frequency. Similarly, we invite students to provide information about their interactions with professors, knowing that we can later remove those interactions if we decide to focus solely on the peer network. In particular, students are asked to ``...choose from the presented list people from [their] physics class that [they] had a meaningful interaction with {\bf in class} ... even if [they] were not the main person speaking or contributing'' (see Tools in the {\it SNA toolbox} for an example of the SNA survey~\cite{sna-tools} for the complete survey). Students are directed to consider all interactions that took place during the week prior to completing the survey, including interactions with peers outside of their small groups. As mentioned earlier, they are not given written parameters for what counted as a ``meaningful'' interaction, but, when asked, we encourage them to think about interactions related to course-related activities and content. To aid recall of their peers' names, we provide them with a randomized roster of all individuals enrolled in class, together with names of the teaching staff. 

\subsubsection*{\texorpdfstring{$\mathcal{Q}3$}{Q3}: \bf\emph{How can I collect network data?}}

There are multiple ways one can collect social network data: videotaping the course, administering a pen-and-paper survey in class, asking students to complete an on-line survey (either in class or at home), using a course-related forum to track students' interaction, etc. Each of these approaches has its own set of pros and cons. With videos one has access to the entire course, which provides a very rich data set. It allows for a fine grain analysis of, e.g., the network evolution in real-time. However, the extraction of networks from videos can be challenging. From establishing a reliable coding dictionary that minimizes coder bias, to determining the most informative time stamp for ``slicing'' the data, to coding what could be hours of videos, this approach requires a lot of time and effort~\cite{Pomian17-SNV}. 

Pen-and-paper surveys take significantly less time, most of which is spent on establishing a protocol for digitizing the responses and converting them into a network. Once established, such protocol can be utilized on consecutive collections. Nevertheless, pen-and-paper surveys require time to develop and place an extra cognitive load on individuals completing them. Moreover, such surveys can be biased and not fully representative of what was happening in class, especially early on when respondents do not know the names of all other participants and relationships are not yet well formed. 

The same applies to online surveys, though in this case converting responses into network data can be handled with a simple script. When administered outside of class time, online surveys tend to suffer from lower response rates. E-mail exchange or forum-based networks offer the same advantages in terms of converting responses into network data with the use of a script. However, as with video data, one has to carefully decide what constitutes an interaction, which is not always straightforward (e.g., handling ``nested'' posts on a forum). Such networks can also suffer from lower response rates, particularly because of missing data from students who read posts or e-mails but do not respond to them~\cite{Traxler18-NFD}. Another thing to consider is whether the participants should receive any incentives for taking part in the study (e.g., course credit, gift cards).

Since a pilot study with both pen-and-paper and computer-assisted versions of the survey revealed that the online approach tends to be more time consuming and more confusing to students, we decided to collect data using the pen-and-paper format. To maximize the response rates we collect data in the classroom, at the end of a particular class. Our participants do not receive any direct benefit from completing the survey (e.g., extra points, reduced workload). Moreover, during the administration students are invited to inquire about the purpose and outcomes of the study by contacting either the professor or the survey administrators.

\subsubsection*{\texorpdfstring{$\mathcal{Q}4$}{Q4}: \bf\emph{How often should I collect network data?}}

Another thing to consider is how often one intends to collect the data and when is the best time for collection. The number of collections should be guided by the research question, collection method, as well as previous research. How much extra burden one is willing to put on students and, for in-class collections, how much class time one is willing to spend on administering surveys also needs to be taken into account. 

In our case, we want to look at students' social embeddedness within the in-class network as a predictor for anxiety shift over time, so it is appropriate to collect network data at least at the beginning and end of the semester. To capture a more granular picture of network evolution, given student-group rotation and other curricular features, we added three additional administrations throughout the semester, spaced every 3-4 weeks. We chose five collections to allow enough time for the network to change between collections. During each survey administration students are reminded that their participation is strictly voluntary. Anecdotal data from past research using similar, in-class surveys suggests that more than five collections may cause survey fatigue. 

\subsubsection*{\texorpdfstring{$\mathcal{Q}5$}{Q5}: \bf\emph{How should I work with longitudinal data?}}

Collecting data multiple times throughout the semester gives one flexibility when preparing data for analysis. Longitudinal data allows for the study of network evolution. Treating each collection as a separate data set enables one to observe changes in the network as time goes by. For instance, comparison of pre- and post-course data from lecture-based and active-engagement classrooms reveals that only in the latter case the in-class network becomes connected, while the former doesn't show any development after a semester of instruction~\cite{Brewe10-PLC}. Analyses of in-class networks from active-learning introductory physics courses show that networks gradually evolve throughout the semester, suggesting that such environments are in fact conducive to establishing a relationship network of academic and emotional support~\cite{Zwolak17-IIP,Williams17-EPP}. However, longitudinal approaches are more sensitive to missingness, as it is quite likely that different individuals may be physically absent during different survey administrations. 

Aggregating multiple collections into one network representing the entire semester helps with missingness, as it is reasonable to assume that each student should be in class during at least one collection. Since the survey distribution schedule was not announced at any point, it seems unlikely that students could intentionally try to avoid classes when data is collected. At the same time, if a given student is absent across multiple survey administrations, it might signal that the individual is skipping more classes and thus is not getting immersed in the social environment. Treating such an individual as disconnected from a classroom network might thus be the appropriate thing to do. However, aggregation limits the amount of information contributing to a complete understanding of the network's evolution~\cite{Dou16-BPM,Zwolak18-PIN}. 

For weighted, directional data there are a multitude of ways network data be aggregated. This can range from simply combining all collections, with weights in the final network calculated as a sum of weights across all collections, to more nuanced computations involving weighted averaging between collections. Alternatively one might simply assign weights based on either the presence or absence of an interaction on a particular collection. The decision of whether to aggregate (and how to proceed with aggregating) should be guided by the research question, previous studies on the population being examined and, if possible, rooted in a theoretical framework. 

Since we ask students to report meaningful interactions that took place during a defined period of time (the week prior to each data collection) and we do so five times during the semester, aggregating all data into one network will result in the loss of information about which interactions happen due to convenience (i.e., sitting at the same table) and which survive the test of time (i.e., recurrent interactions regardless of group membership). Thus, in our analyses we treat each collection as a separate network. This allows us to capture the effect of modifications to the seating arrangements and the group exam on the evolution of the network throughout the semester.

\subsubsection*{\texorpdfstring{$\mathcal{Q}6$}{Q6}: \bf\emph{How can I quantify social interactions?}}

Some of the remaining considerations include how to convert interaction data into a network and then how to analyze the resulting network. As mentioned when discussing the different tools for collecting SNA data (Q3), the protocol for converting data into a social network will depend on the particular data collection approach. When digitizing data, one should retain the capability of formatting identified interactions as interaction matrices or lists of the pairs involved in an interaction (i.e., edge lists). Once a matrix or an edge list is created, SNA provides a very rich toolbox for analysis. From various network topology measures to a multitude of centralities, there is plenty to choose from. In general, one can examine the interactions in a network from one of two broad perspectives: whole network connectedness (i.e., network topology) and individual node-level measures (i.e., centralities). 
 
To digitize our pen-and-paper surveys into networks, we developed a spreadsheet with built-in self-checks in order to minimize coding errors. The spreadsheet is available as part of the {\it SNA toolbox}~\cite{sna-tools}. As mentioned earlier, we opted to keep each collection as a separate network. To examine students' interactions, we calculate three centrality measures discussed in Sec.~\ref{sub:sna-tools}. Our choice of these particular indices is guided by their ability to capture the kind of immersion within the network that we hypothesize to be relevant for anxiety shifts --  overall embeddedness in the case of closeness and individual-level connectedness in the case of indegree and outdegree. This approach is also supported by previous research that found these measures to be informative when studying performance~\cite{Williams17-EPP} and persistence~\cite{Zwolak17-IIP,Zwolak18-PIN}, both of which are related to anxiety.

\subsection{Social Network Analysis toolbox}\label{sub:sna-tools}
There are two basic types of static network measures: the network-level measures that describe characteristics of the network as a whole and the node-level measures that focus on characterizing the relational position of a particular node quantitatively. In what follows, we use the term ``node'' in reference to the individuals that make up a social network (note that social sciences often use the term ``actor'' instead) and ``edge'' (also called ``tie'' or ``link'') when referring to the interaction between two nodes. The following section gives a brief overview of the most commonly used tools for quantifying interactions from an SNA perspective. All metrics discussed below are implemented in the {\it SNA toolbox}~\cite{sna-tools}. 

When choosing to combine data across multiple groups (e.g., multiple sections of the same course), it is important to verify that the networks are similar enough to justify aggregation. Network topology offers understanding of how nodes are connected with one-another on a global level. This includes characteristics like network size, density, and distances between nodes. For example, {\it density} ($\Delta$) offers insight about the overall cohesion of a network and is expressed as the fraction of existing edges between nodes to the number of all possible edges:
\begin{equation*}
\Delta=\frac{number\,of\,present\,\,edges}{number\,of\,all\,\,possible\,\,edges}.
\end{equation*}
The number of all possible edges between $n$ nodes is expressed as $n(n-1)/2$ for undirected graphs and as $n(n-1)$ for directed graphs~\cite{Wasserman1994}. Density analyses produce values between $0$ and $1$. Active-learning physics classrooms have been shown to exhibit greater density than traditional, lecture-based classrooms~\cite{Brewe10-PLC}. 

Network diameter and average path length are other metrics related to network-level connectedness. {\it Diameter} ($D$) gives a network's longest path -- where path is defined as the number of edges in the sequence of edges connecting two nodes in a network -- and captures the span of a network. {\it Average path length} ($L$), on the other hand, gives the average shortest path between all possible pairs of nodes. It provides information about how close (on average) nodes are to one another~\cite{Wasserman1994}. 

The {\it global clustering coefficient} ({\it transitivity}, Tr) captures the degree to which nodes tend to cluster together. It is based on the notion of open and closed triplets in a network, where a triplet is defined as three nodes connected by either two (open triplet) or three (closed triplet) undirected edges~\cite{Wasserman1994}. Transitivity is defined as a fraction of closed triplets of all triplets (opened and closed) in the network:
\begin{equation*}
\text{Tr}=\frac{number\,of\,closed\,triplets}{number\,of\,all\,triplets},
\end{equation*}
Since by definition transitivity is calculated for undirected and unweighted networks, networks that are more complex in nature have to be flattened prior to analysis. This, in return, allows one to vary the strength of transitivity. For instance, requiring that all edges in triplets are bidirectional will lead to a stronger global clustering coefficient than the simple presence or absence of edges. Similarly, requiring that all edges in a triplet are of weight at least $w$, where $w\ge1$, will result in stronger transitivity the larger $w$ is. Recently, a generalization of the global clustering coefficient that includes weight was proposed~\cite{Opsahl09-CWN}. Since we use transitivity only to establish similarity between our networks and do not use it in analysis, we find the basic, binary version to be sufficient.

Finally, {\it reciprocity} captures how frequently interactions are mutual. It is calculated as a fraction of all the interactions that are bidirectional~\cite{Wasserman1994}:
\begin{equation*}
\rho^{\leftrightarrow}=\frac{number\,of\,bidirectional\,\,edges}{number\,of\,all\,\,present\,\,edges}.
\end{equation*}

\begin{figure}[t]
\centering
\includegraphics[width=1.0\linewidth]{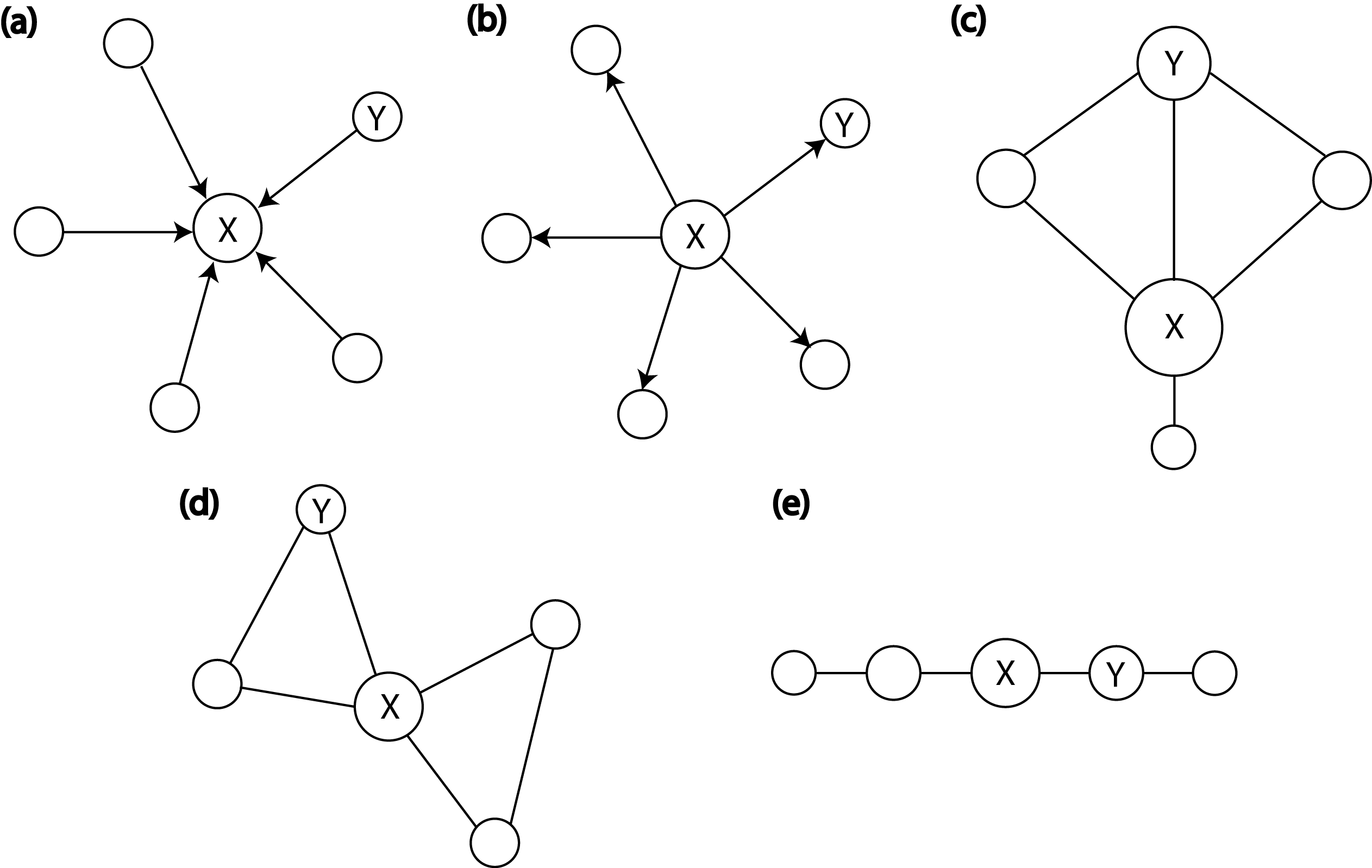}
\caption{Visualization of various types of centralities. In each case, X has higher centrality than Y according to (a) indegree, (b) outdegree, (c) eigenvector, (d) closeness, and (e) betweenness. Adapted with permission from Ref.~\cite{Zwolak17-IIP}.\label{fig:centralities}}
\end{figure}

Once similarity of networks between groups is confirmed, one can proceed to quantifying the position of each node within the network. This is most commonly done by calculating centrality measures. There are a myriad of such measures, from localized, i.e., focused on a particular node and its direct connections (see, e.g., Fig.~\ref{fig:centralities}(a) and (b)), to global measures that take into account the entire network (see, e.g., Fig.~\ref{fig:centralities}(d) and (e)). The choice of a particular measure depends of the context of the study. There are various textbooks that give a good introductory~\cite{Prell2011} and more advanced~\cite{Wasserman1994,Scott2011} overview of centrality measures, as well as primers that explain their use in different contexts (see, e.g., Ref.~\cite{Grunspan14-SNA} for primer in education research). Here, we only briefly describe measures that we use in our analysis. 

Building on our previous work~\cite{Dou16-BPM,Zwolak17-IIP,Williams17-EPP,Dou18-USE,Zwolak18-PIN}, we calculate the following three measures: indegree, outdegree and closeness. Put simply {\it indegree} can be thought of as a measure of popularity. It is calculated as the number of edges directed towards a given node. {\it Outdegree} --  the number of edges that a given node sends to others -- can be interpreted as sociability or influence. Finally, {\it closeness} captures how well a given node is embedded within the entire network -- the ``closer'' a given node is to everyone else in the network, the more access that person might have to resources (e.g., knowledge, educational or emotional support, information about study groups). Here we use the weighted generalization of these measures that accounts for both the edges' weights and their number~\cite{Opsahl10-CWN}, with the parameter $\alpha$ tuning the relative importance of these two factors. Formally, for degree
\begin{equation}\label{eq:degree}
[C_{D}^{\alpha}](i)=(i's\,\,binary\,\,degree) \left[\frac{i's\,\,strength}{i's\,\,binary\,\,degree}\right]^\alpha,
\end{equation}
where $\alpha\in[0,\infty)$ is the tuning parameter, the node's binary degree is the number of incoming edges for indegree and outgoing edges for outdegree, and the node's strength is a sum of weights of incoming edges for indegree and outgoing edges for outdegree. If $\alpha=0$, then $C_{D}^{\alpha}$ gives the binary degree and if $\alpha=1$, then $C_{D}^{\alpha}$ returns the overall sum of all weights (i.e., strength). When $\alpha\in(0,1)$, having many weak connections is emphasized over a few strong ones (keeping overall strength fixed). When $\alpha>1$, it is favorable to have a few strong connections (for the same total strength). 

For closeness, 
\begin{equation}\label{eq:close}
C_{C}^{\alpha}(i)=\left[\begin{smallmatrix} sum\,of\,weighted\,shortest \\ paths\,to\,all\,other\,nodes \end{smallmatrix}\right]^{-1},
\end{equation}
where the weighted path linking $i$ and $j$ is defined as $d_{ij}^{\alpha}=\min\left(w_{im}^{-\alpha}+\dots+w_{nj}^{-\alpha}\right)$. Like with degree, for $\alpha=0$, the binary version of closeness results (i.e., the weights are ignored), while for $\alpha=1$ only the weights are important. If $\alpha\in(0,1)$, a shorter path of weak ties is favored over a longer path with strong ties and for $\alpha>1$ the number of intermediary nodes is less important than the strength of the ties. To explore the relative importance of the number of ties and their weights we use multiple values for the $alpha$ coefficient. 

\subsection{Other considerations}\label{sub:other}
\subsubsection{Accounting for non-normality}\label{sub:non-norm}

Given the interdependence of network data, its distribution often fails tests of normality. For example, when student A reports one outgoing interaction with peer B, by definition a researcher records an incoming interaction for peer B. Because one student's responses can affect another student's responses, interaction data often violates the assumption of independence required by typical statistical analyses. Moreover, centrality measures are not always normally distributed, which violates requirements of linear regression models.

To account for these violations, we use linear regression permutation tests~\cite{Anderson01-PTL}. Linear regression permutation tests use a type of Monte Carlo method to randomly sample a data set, rearranging the values of its variables across all observations. A linear model is tested on this re-sampled data set, which generates a set of regression estimates. The regression estimates of the original data set are then compared to the distribution of estimates generated from the permuted sets in order to determine the reliability of the outcomes. In addition to not requiring data to be normally distributed, this kind of test helps to minimize the false positive finding (i.e., type I error). 

\subsubsection{Handling missing network data}\label{sub:missing_data}

Regardless of whether the data collection takes place in or outside of the classroom, through pen-and-paper or on-line surveys, it is quite unlikely that any given collection will solicit a 100~\% response rate. Students may not show up to class on a day when data is collected, they might leave early, or may choose not to complete the questionnaire. In any case, response rates should be considered when choosing an approach for handling missing data. To do so, one must first define the network boundaries.

Classroom networks can be defined by one of two typical boundaries: (A) students officially enrolled in the class or (B)  students who choose to share network data. The former treats all enrolled students as members of a network on each collection, with absentees and non-respondents contributing to the overall ``missingness'' of the network. The latter boundary posits classroom participation (e.g., attendance on the day of data collection) as a qualifier for inclusion in the network. Both approaches have pros and cons. Boundary ``A'' is most inclusive, taking into account the behavior of all students enrolled in a course, regardless of their attendance or participation throughout the semester. Research questions that aim to understand broad ranges of social behaviors lend themselves to this approach. On the other hand, researchers interested in specific-types of behavior (e.g., peer-peer interactions) may want to take the second approach and limit the network boundary to those present, given that a student's absence does not necessarily reflect their in-class social behavior. Regardless of the approach missingness will almost always be present.

The challenges that result from missingness in a network stem from the inherent interdependence of network data. A student's behavior in a network not only affects their position in the network, but also the position of others in the network regardless of whether or not the student in question directly interacts with everyone in the network. Typical methods for handling missing data, such as imputation techniques, do not take into account data interdependency; while they may predict a given individual's centrality scores, they fail to account for how that would affect the scores of all others in the network. Replacing missing data with substitute values increases the chances of significantly changing the properties of the network. On the other hand, it has been shown that centrality scores are fairly robust to random missingness. For example, for small networks ($40$ -- $75$ nodes) the level of missing data that does not affect the overall structure is up to $35\%$ for directed degrees and about $20\%$ for closeness centrality~\cite{Smith13-SES}. The missingness in our network data falls within these thresholds and therefore no imputation was used. However, if the missingness falls outside of those thresholds, it may not be possible to do a whole-network analysis. One can still try to examine ego-networks, i.e., build networks based on all data available but look only at individuals who responded to the survey. Such initial analysis can be then complemented by, e.g., interviews or data from registrars. In either case, caution should be taken when drawing conclusions in light of what data is available.
 
\subsection{Statistical analysis}\label{sub:stats}
The dependent variable in our study is continuous (the normalized shift in anxiety). To investigate relationships between students' pre-course anxiety, network centralities, gender, final grade and their shift in physics anxiety, linear regression modeling is used. To control for confounding factors, we perform multiple linear regression. Only significant variables for the simple linear regression analysis are incorporated into the full model.   

In the first stage, we want to determine which centralities carry significant information about the anxiety shift. To do so, we run simple linear regression models with a single centrality as a predictor (i.e., $anxiety.shift \sim centrality$). To explore the relative effect of the number of edges and their weights, we test four values of the tuning coefficient: $\alpha=0.0$ (only the number of edges matters, weights are ignored), $\alpha=0.5$ (it is better the have more edges, keeping strength fixed), $\alpha=1.0$ (only the total strength matter, regardless of the number of edges) and $\alpha=1.25$ (it is better to have less edges, keeping strength fixed), see Sec.~\ref{sub:sna-tools} for details. 

In the second phase, we want to take advantage of the longitudinal nature of our data. Having identified the statistically significant centralities from the last survey administration, i.e., our fifth collection, we investigate which of those measures remain significant on earlier administrations. To do so, we test simple linear regression models for all earlier collections, i.e., collections one through four. We then compare the fits of the models to determine the relative importance of the number and weights of edges and identify the most useful tuning parameter $\alpha$ value for our purposes.

Finally, after identifying the earliest informative collection and $\alpha$ value, we move to testing full linear models (i.e., $anxiety.shift \sim centrality + gender + final.grade + pre.anxiety$). The variance inflation factor for the final model, ranging from 1.0 to 1.1, indicates no collinearity between variables.
 
To account for the fairly large number of tested models, we run each test as a permutation test. As previously described, permutation test randomizes the matching between independent and dependent variables and compares the true regression estimates to the distribution of estimates calculated across a certain number of iterations of randomization. In our study, we use 5000 iterations. Again, the use of permutation tests helps to address two concerns that arise when dealing with network data: (1) missing data and (2) violation of the assumptions of normality and homoscedasticity (i.e., same finite variance for all random variables in the sequence).

For the statistical analyses, we use the R statistical programming language~\cite{R}. In particular, we use lmPerm~\cite{lmPerm,Wheeler16-PTR} package for the permutation test for linear models, the Amelia~\cite{amelia} package for imputation of anxiety data, and the igraph~\cite{igraph} and tnet~\cite{tnet} packages for network analysis. The chi-squared test and Fisher's exact test are used to test for statistically significant differences between classroom sections in terms of gender and ethnicity. The one-way analysis of variance (ANOVA) is used to compare the two section in terms of students' GPA and paired t-test is used to compare the anxiety scores between sections. The Kolmogorov-Smirnov test is used to compare the original and imputed PARS scores, and Shapiro-Wilk test is used to test for normality of the centrality scores' distributions. To adjust the false discovery rate the Benjamini-Hochberg procedure is implemented \cite{Benjamini95-CFD}. We consider results with $p<.05$ as significant. All protocols in the project were approved by the Florida International University Institutional Review Board (IRB-13-0240 exempt, category 2).

\section{Analysis \& Results}\label{sec:analysis}
This section describes practical applications of the proposed methodologies in the context of students' physics anxiety in introductory physics courses. We set out to understand whether students' social interactions and positioning in the classroom network is predictive of their shift in anxiety while controlling for their pre-course anxiety, self-reported gender and final course grade. We also want to understand when during the semester, if at all, does social integration begin to matter with regard to shifts in anxiety.

\subsection{Demographics}\label{sub:demog}
The data for this study was collected at a large research university, designated as a Hispanic-Serving Institution. In particular, we survey students enrolled in the Introductory Physics I with Calculus course taught using the Modeling Instruction (MI) curriculum. Due to its inquiry-laden, discourse-based approach, MI provides an ideal context for studying the range of possible student-student interactions in an introductory physics classroom~\cite{Hestenes87-MIP,Halloun87-MIM}. The course combines lab and lecture components of Physics~I, engaging students with hands-on, group activities in which they develop models of physical phenomena through the use of various representations (e.g., equations, graphs, diagrams or a combination thereof)~\cite{McPadden17-MIR}. Students work in small groups of three, with two small groups typically sharing a table, in order to develop representations relevant to the problem at hand. Then students come together in larger groups of about 25 to 30 to discuss the small group findings. Instructors, teaching assistants, and learning assistants facilitate both large and small group discussions. Traditional lecture rarely occurs during the semester. Instead, students participate in a flexible classroom space designed for active-learning. Chairs and tables are movable and students are provided with portable white boards. They are permitted to communicate with peers in other groups and often do so. Small group membership is randomly selected and changes several times throughout the semester.

\begin{table}[t]
\renewcommand{\arraystretch}{1.1}
\renewcommand{\tabcolsep}{6pt}
\caption{Students' gender and ethnicity distribution. The numbers represent the percentage of students is given group.}
\centering
\begin{tabular}{p{2.5cm}p{1.5cm}p{1.5cm}} \hline \hline
  & Section A &  Section B   \\ \hline  
Gender: Female    &  41.5 & 50.7 \\ \hline
Ethnicity &  &   \\   
\hspace{3mm}Asian     & 15.1 & \hspace{1.5mm}2.7 \\
\hspace{3mm}Black     & 11.3 & 12.3 \\
\hspace{3mm}Hispanic  & 60.4 & 72.6 \\
\hspace{3mm}White     & \hspace{1.5mm}3.8 & \hspace{1.5mm}8.2 \\
\hspace{3mm}Other/NA  & \hspace{1.5mm}9.4 & \hspace{1.5mm}4.1\\
\hline \hline
\label{tab:demog}
\end{tabular}
\end{table}

\begin{figure*}[t]
\centering
\hfill
\begin{minipage}{0.45\linewidth}
\raggedright
(a)\\
\centering
\includegraphics[width=\linewidth]{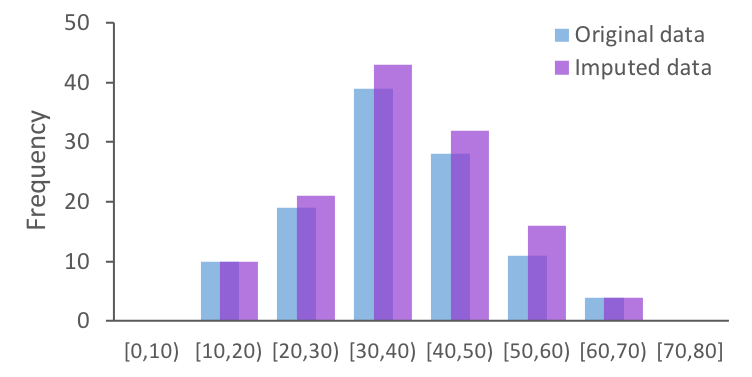}
\end{minipage}
\hfill
\begin{minipage}{0.45\linewidth}
\raggedright
(b)\\
\centering
\includegraphics[width=\linewidth]{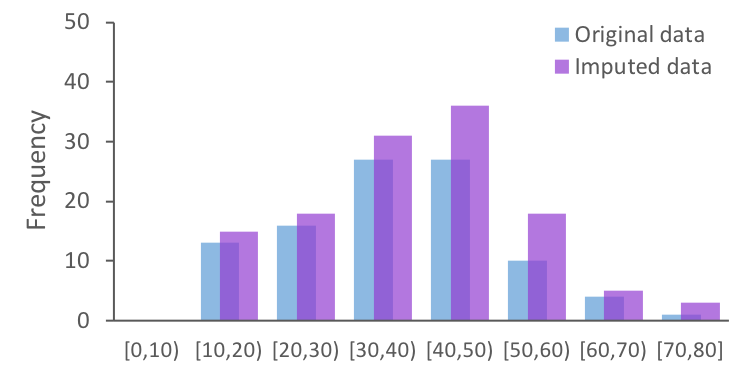}
\end{minipage}
\hfill
\caption{Comparison of the original and imputed data for (a) the anxiety pre score and (b) the anxiety post score. \label{fig:anx-dist}}
\end{figure*}

The data for this analysis comes from two MI sections offered in fall 2016 ($N_{F16A}=53$, $N_{F16B}=73$). There were two instructors teaching the course, both with several years of experience teaching introductory physics using student-centred curricula, including MI. Student demographic data was queried from a university database and includes self-reported gender (binary: female or male), incoming GPA, and final course grade, see Table~\ref{tab:demog} for details. We find no statistically significant differences between sections in terms of gender (chi-squared test, $\chi^2(1)=0.70$, $p=0.40$) and ethnicity (Fisher's exact test, $p=0.06$) distributions. There is also no significant difference in mean incoming GPA between groups (one-way ANOVA, $F(1,123)=2.04$, $p=0.16$, note that the GPA for one student was not available).

\subsection{Analysis of physics anxiety}
Students' total scores on the PARS were generated by adding up the sum of their scores on the individual items on the survey. Paired samples $t$-tests showed no significant difference between the mean {\it pre} and {\it post} physics anxiety total scores, regardless of instructor ($t=0.74$, $p=0.46$ for Section A, $t=-1.74$, $p=0.09$ for Section B) , nor when combining instructor data ($t=-0.65$, $p=0.52$). Since not all students were present when the anxiety survey was administered, there were missing scores: 8 for {\it pre} survey, 21 for {\it post} survey, and additional 7 for both. To account for the missing data, we ran a single imputation. Figures~\ref{fig:anx-dist}(a) and~\ref{fig:anx-dist}(b) show the comparison of distribution for the {\it pre} and {\it post} scores for original (blue) and imputed (purple) data, respectively. The two sample Kolmogorov-Smirnov test showed no statistically significant differences in the distributions, with $p=1$ for both {\it pre} and {\it post} scores.

\begin{figure}[b]
\raggedright
(a)\\
\includegraphics[width=\linewidth]{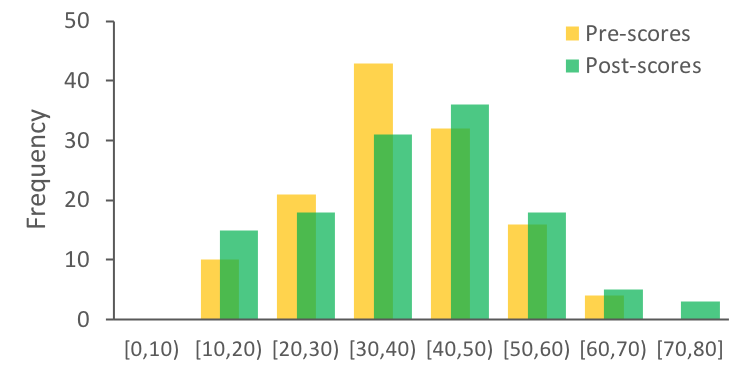} \\
\raggedright
(b)\\
\includegraphics[width=\linewidth]{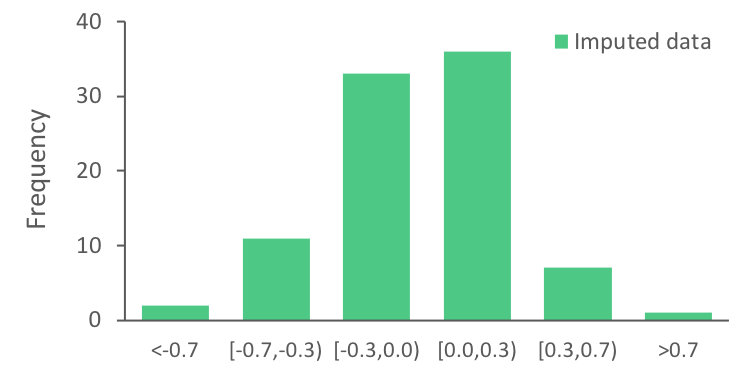}
\caption{(a) The distribution of the imputed {\it pre} (yellow) and {\it post} (green) anxiety scores showing slight right skewing. (b) The normalized shift at individual level as defined in Eq.~(\ref{eq:nor-shift}).
\label{fig:anx-shift}}
\end{figure}

With the imputed data, the average anxiety score at the beginning of the semester for instructor A's section was $M_A^{pre}=38.2$ (standard deviation $SD_A^{pre}=11.3$) and for instructor B's section -- $M_B^{pre}=37.5$ ($SD_B^{pre}=12.2$). The average anxiety measure at the end of the semester in instructor A's class was $M_A^{post}=41.9$ ($SD_A^{post}=13.3$) and in instructor B's class $M_B^{post}=36.2$ ($SD_B^{post}=13.5$). Given the lack of statistically significant differences between the two instructors ($t$-test, $t=0.89$, $p=0.38$) we combined the data from their courses ($N$=126). 

The range of the imputed PARS scores went from $(16,63)$ at beginning of the semester to $(16,79)$ at the end (for the non-imputed {\it post} scores it is $(16,74)$). The range increases slightly from {\it pre} to {\it post} responses. Qualitative analysis of histograms reveals slight right skewing when comparing the scores from {\it pre} to {\it post}, indicating that while the overall mean did not change significantly across the semester, individual students' anxiety did experience some shifts, see Fig.~\ref{fig:anx-shift}(a). For the following analyses, we use individual students' normalized shift in anxiety in order to take into account their maximum possible shift, see Fig.~\ref{fig:anx-shift}(b) for the shift's distribution. 

\begin{figure*}[t]
\centering
\hfill
\begin{minipage}{0.32\linewidth}
\raggedright
(a) First collection\\
\centering
\includegraphics[width=\linewidth]{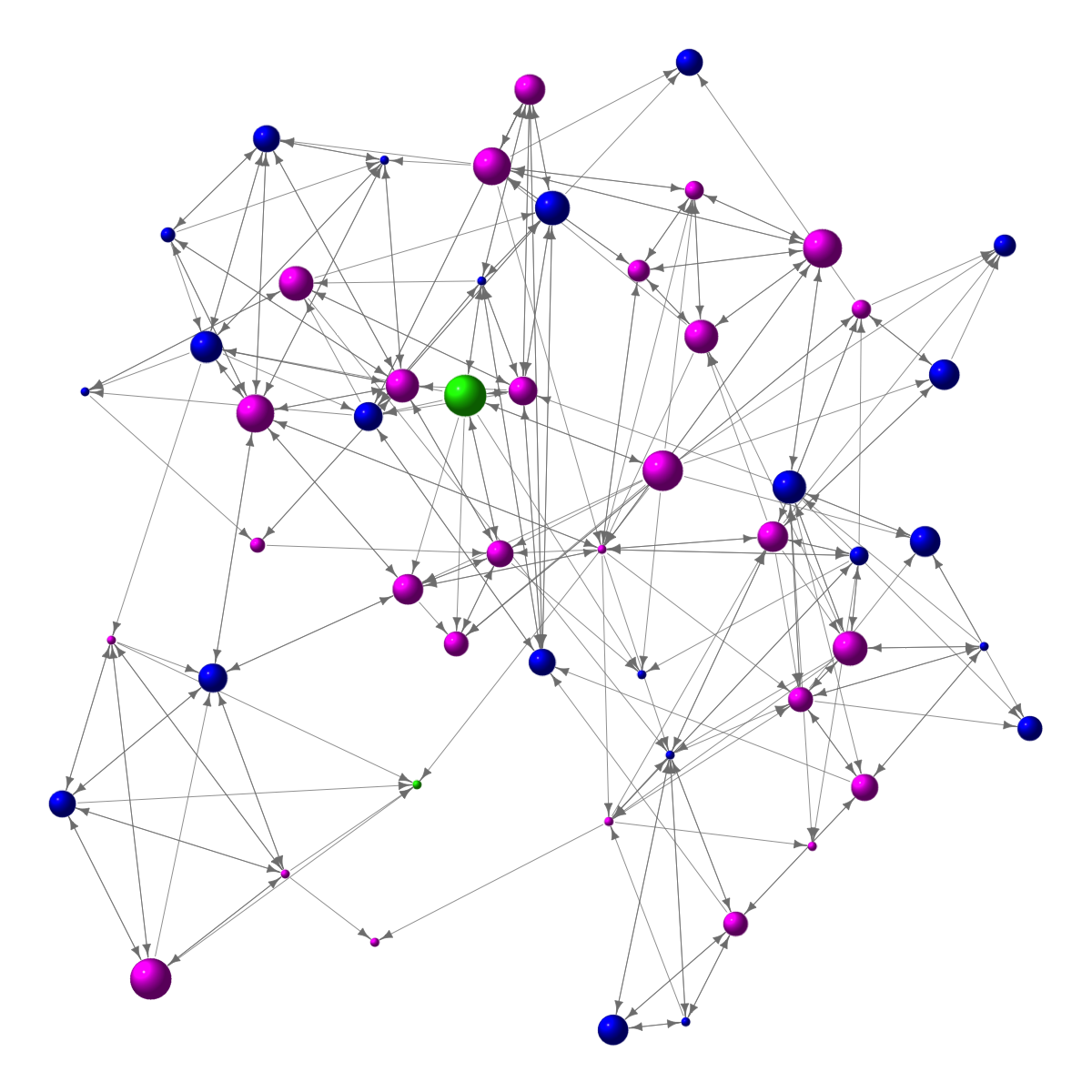}
\end{minipage}
\hfill
\begin{minipage}{0.32\linewidth}
\raggedright
(b) Fourth collection\\
\centering
\includegraphics[width=\linewidth]{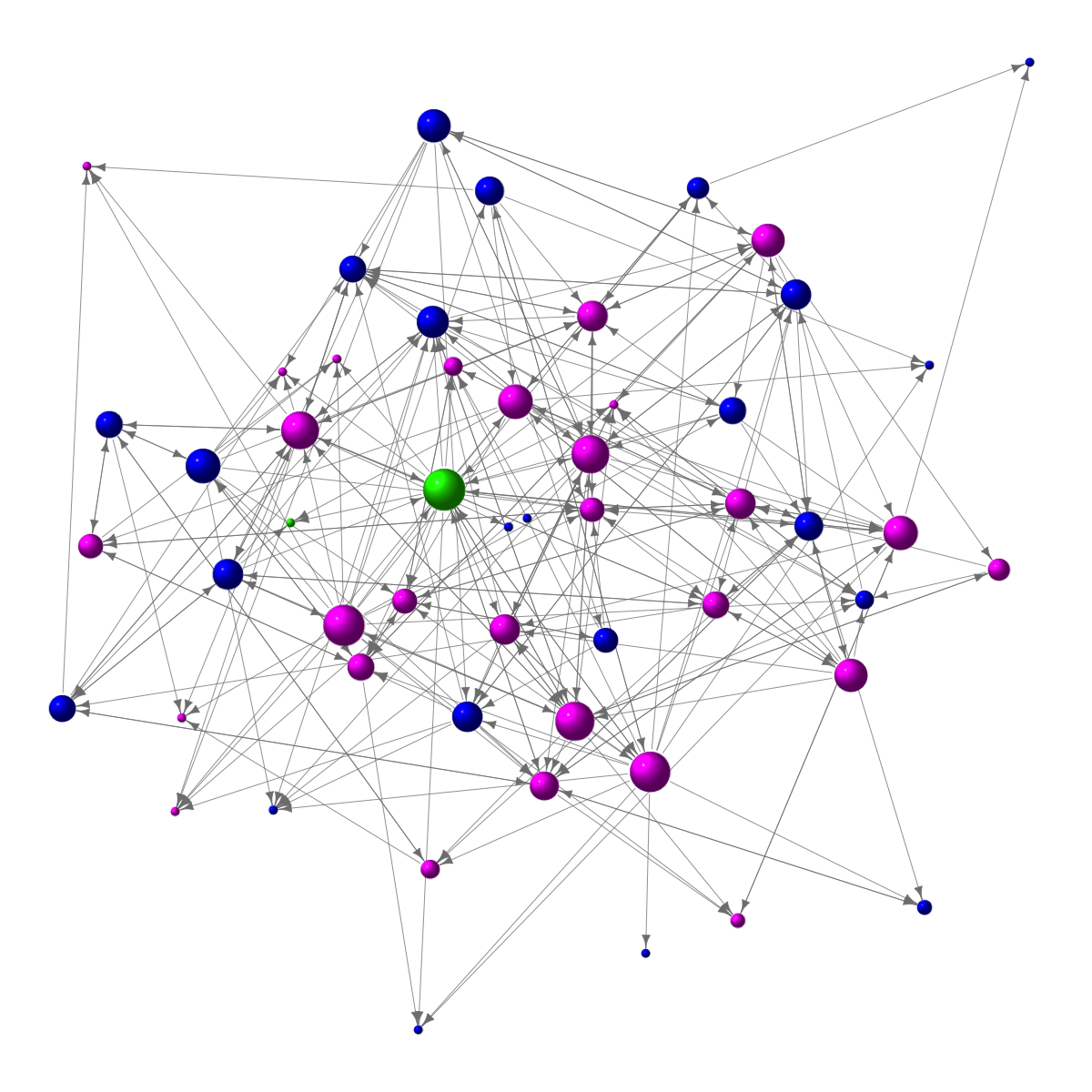}
\end{minipage}
\hfill
\begin{minipage}{0.32\linewidth}
\raggedright
(c) Fifth collection\\
\centering
\includegraphics[width=\linewidth]{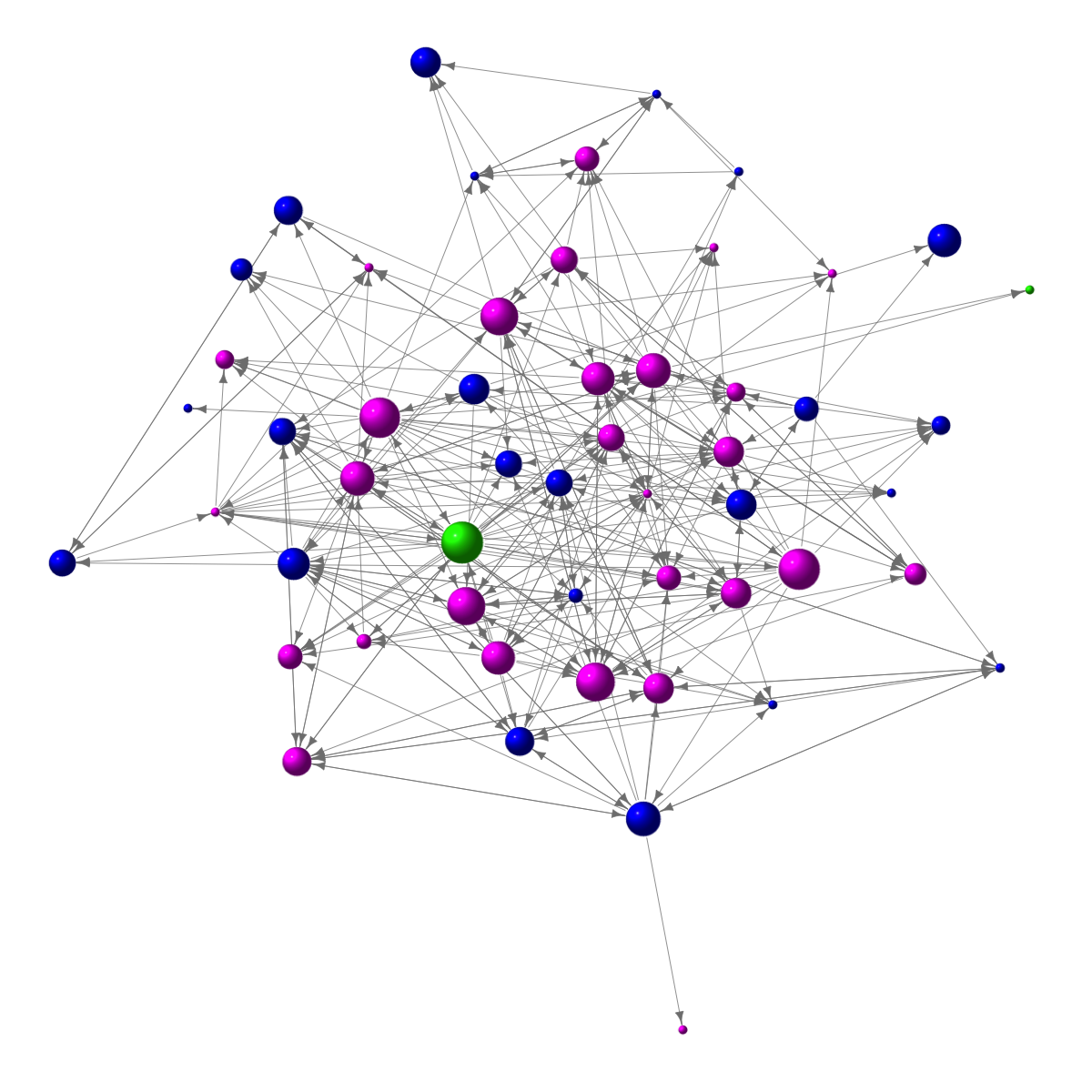}
\end{minipage}
\hfill
\caption{The in-class network evolution for individual networks in Section A ($N=53$), with the size representing the outdegree from fourth collection ($\alpha=0.0$) and the color indicating the direction of the shift in anxiety scores (magenta -- positive, blue -- negative, green -- no shift. The networks include only students.\label{fig:net-evol}}
\end{figure*}

\subsection{Analysis of student networks}
As mentioned in Sec.~\ref{sub:sna-tools}, when analyzing network data from multiple groups, it is important to verify that there is foundation for aggregating the data. The response rates to the survey were fairly comparable between sections: 
$M_{A}=80.2$ ($SD_{A}=6.8$) and $M_{B}=79.4$ ($SD_{B}=11.0$). The Kruskal-Wallis test shows no statistically significant differences in response rates between the two sections ($\chi^2(1)=0.01$, $p=0.92$). The whole network characteristics, as well as various students' centrality scores, were calculated separately for each section. Table~\ref{tab:net_info} shows the comparison of network characteristics for the two sections at first, fourth, and fifth collections. 

\begin{table}[b]
\renewcommand{\arraystretch}{1.1}
\renewcommand{\tabcolsep}{9pt}
\caption{The comparison of network characteristics for first (SNA1), fourth (SNA4), and last (SNA5) collection for fall 2016 (two sections, A and B): network size ($n$), density ($\Delta$), average path length ($L$), diameter ($D$), transitivity ($Tr$) and reciprocity ($\rho^{\leftrightarrow}$). Note that instructional staff is removed from the network.}
\centering
\begin{tabular}{l|cccccc} \hline \hline
  & $n$ & $\Delta$  & $D$ & $L$ & $Tr$ & $\rho^{\leftrightarrow}$ \\ \hline
SNA1 A & 53 & $0.10$ & $7$ & $3.0$ & $0.46$ & $0.69$\\
SNA1 B & 73 & $0.09$ & $8$ & $3.2$ & $0.39$ & $0.65$\\ \hline
SNA4 A & 53 & $0.11$ & $5$ & $2.4$ & $0.25$ & $0.44$\\
SNA4 B & 73 & $0.07$ & $7$ & $2.9$ & $0.29$ & $0.45$\\ \hline
SNA5 A & 53 & $0.13$ & $6$ & $2.5$ & $0.33$ & $0.42$\\
SNA5 B & 73 & $0.09$ & $6$ & $2.7$ & $0.34$ & $0.52$\\ 
\hline \hline
\end{tabular}
\label{tab:net_info}
\end{table}

As can be seen in Table~\ref{tab:net_info}, the networks have fairly comparable topologies and patterns of interactions, with the network in Section A being slightly denser and with a somewhat smaller diameter, which is to be expected of a smaller network. Visualization of networks generated from first, fourth, and fifth collections are shown in Fig.~\ref{fig:net-evol} and the descriptive statistics for centralities are presented in Table~\ref{tab:stats_centr_4} and Table~\ref{tab:stats_centr_5} in Appendix~\ref{app:desc_centr}.

\subsection{Predicting shifts in anxiety}
\begin{table*}[t]
\renewcommand{\arraystretch}{1.1}
\renewcommand{\tabcolsep}{10pt}
\caption{Summary of the linear regression for anxiety shift as predicted by weighed outdegree from fourth and fifth collection, with $\alpha\in\{0.0, 0.5, 1.0, 1.25\}$: the unstandardized estimate (B), the standard error for the unstandardized estimate (SE B), standardized estimate ($\beta$), $t$-test statistic ($t$), and R-squared ($R^2$). We consider networks without instructional staff. Significant $p$-values are marked with an asterisk.}
\begin{tabular}{c|ccccc|ccccc} \hline\hline
\multirow{2}{*}{Centrality} & \multicolumn{5}{c|}{Fourth collection} & \multicolumn{5}{c}{Fifth collection} \\
& B & SE B & $\beta$ & $t$ & $R^2$ & B & SE B & $\beta$ & $t$ & $R^2$\\ \hline
$C_{outD}^{0.0}$ & $-0.02^{**}$ & $0.007$ & $-0.28^{**}$ & $-3.28$ & $0.08$
                 & $-0.02^{*}$ & $0.007$ & $-0.22^{*}$ & $-2.51$ & $0.05$\\ 
$C_{outD}^{0.5}$ & $-0.02^{**}$ & $0.005$ & $-0.28^{**}$ & $-3.28$ & $0.08$
                 & $-0.01^{**}$ & $0.005$ & $-0.25^{**}$ & $-2.84$ & $0.06$\\ 
$C_{outD}^{1.0}$ & $-0.01^{**}$ & $0.004$ & $-0.28^{**}$ & $-3.22$ & $0.08$
                 & $-0.01^{**}$ & $0.003$ & $-0.27^{**}$ & $-3.16$ & $0.07$\\
$C_{outD}^{1.25}$& $-0.01^{**}$ & $0.003$ & $-0.27^{**}$ & $-3.16$ & $0.07$
                 & $-0.01^{**}$ & $0.002$ & $-0.28^{**}$ & $-3.30$ & $0.08$\\ 
\hline \hline
\multicolumn{7}{l}{\footnotesize **$p<0.01$, *$p<0.05$}
\end{tabular}
\label{tab:centr4_5-outD}
\end{table*} 

Depending on the number of data collections that best fits a study, as well as the number of constructs being explored, the number of variables that need to be considered can become enormously large and the number of statistical tests to run can reach values that make false positive findings more likely. Eliminating irrelevant variables helps to ameliorate some of these concerns. With the abundance of various centrality measures, each having its own advantages and disadvantages, and usually quite different interpretations, it might seem appealing to try as many as possible and ``see what works''. However, as we stressed earlier, the choice of particular metrics should be made in light of previous research whenever possible. 

In our case, prior studies indicate that students' networks in an active-learning classroom evolve over time, and that in the case of persistence and academic performance in physics, social networks established by about half way through the semester become more informative~\cite{Williams17-EPP,Zwolak18-PIN}. With regard to physics anxiety, however, we found no study that explores it in the context of students' classroom network evolution. For this reason we choose to begin our exploration with student networks at the end of the semester, i.e., from the fifth SNA survey administration. At this point in the semester students have had ample opportunity to interact with nearly all of their classmates, either in small groups, board meetings, or one-on-one. Moreover, multiple rotations of seating assignments facilitated and encouraged more extensive interacting through, e.g., team work, labs, and other assignments with different groups of students. Therefore, students had the greatest amount of information with which to evaluate the level and quality of their interaction with classmates within and outside of their small groups. 

In order to further reduce the number of variables, we employ a four phase approach in such a way that each subsequent phase of analyses takes into account a narrower, but more relevant set of factors. 

\subsubsection*{\bf\emph{Phase I: Which centrality measures contribute to anxiety shift?}}

Given our exploratory approach to investigating the relationship between students' embeddedness within the in-class network and anxiety, we run simple linear models looking at the predictive value of the centrality indices presented in Sec.~\ref{sub:sna-tools} on the normalized anxiety shift. The simple models test three measures of centrality as independent variables: indegree, outdegree, and closeness. Because it is unclear from the perspective of physics anxiety whether it is more important to weigh repeated interactions with the same individuals as opposed to multiple interactions with different individuals, we calculate each centrality measure using four different tuning parameters $\alpha$~\cite{Opsahl10-CWN}. As discussed in Sec.~\ref{sub:sna-tools}, $\alpha$ allows to control for the relative importance of the number of edges and their weights (see Eq.(\ref{eq:degree}) and Eq.(\ref{eq:close})). The four values we choose, $\alpha\in\{0.0, 0.5, 1.0, 1.25\}$, reflect four different ways to weigh the strength of repeated interactions between the same two individuals. In what follow, we use the subscript convention to indicate which centrality we refer to (i.e., $inD$ for indegree, $outD$ for outdegree and $C$ for closeness) and superscript for the tuning parameter used to weigh interactions when calculating a particular type of centrality measure (e.g., $C_{inD}^{1.0}$ denotes indegree with $\alpha=1.0$).

We run a simple linear regression for each centrality measure calculated using the tuning parameters listed above, i.e., $M_{slr} : anxiety.shift \sim centrality$. This gives 12 different tests, four for each measure. Each test is run as a permutation test for linear models to verify its statistical significance. Our tests on the network data collected at the end of the semester reveals no significant relationship between normalized anxiety shifts and indegree, regardless of the tuning parameter value. Outdegree (regardless of $\alpha$) and closeness ($\alpha>0$) are significant predictors of normalized anxiety shift. However, when adjusted for false positives (type I error), only outdegree remains significant (for all $\alpha$). The negative estimates suggest that the greater a students' outdegree, the more likely that student is to experience a larger decrease in their anxiety from the beginning to the end of the semester (see Table~\ref{tab:centr4_5-outD} for the regression estimates for outdegree from fifth collection). The standardized beta estimates $\beta$ range from $-0.22$ to $-0.28$, with an average of $-0.26$. In other words, on average, for every one standard deviation increase in a student's outdegree, their normalized physics anxiety would decrease by $0.26$ standard deviation. This shift could be characterized as either negative, as compared to anxiety at the beginning of the semester, or simply a decrease compared to other students but still positive compared to anxiety at the beginning of the semester.

\subsubsection*{\bf\emph{Phase II: When do centrality measures start to matter?}}

In order to implement an intervention aimed at mitigating students' physics anxiety, it is important to know which students are ``at risk'' when there is still time to intervene. Thus, we seek to identify when during the semester might be an appropriate time to do so. Since we have access to data collected five times throughout the semester, we proceed to investigate the correlation between anxiety shift and outdegree on earlier collections. We run simple linear regressions with outdegree as a predictor of normalized anxiety shifts, i.e, $M_{slr} : anxiety.shift \sim centrality$, for each of the four untested data sets, i.e., collections one through four. We test each collection for the same values of the tuning parameter $\alpha$ as in Phase I. These tests are also run using permutation techniques. We find outdegree to be a significant predictor of normalized anxiety shift beginning in collection four, regardless of the tuning parameter used (see Table~\ref{tab:centr4_5-outD} for the regression estimates for outdegree from fourth collection). Outdegree is not significantly correlated with the shift in physics anxiety for collections one, two, and three. 

\subsubsection*{\bf\emph{Phase III: Which tuning parameter makes the most sense?}}

The tests described in Phase II reveal that outdegree centrality begins to play a role in students' physics anxiety shift sometime around the fourth data collection, which took place after the second midterm which also happens to be a group exam. In order to determine how to best weigh repeated interactions between the same two individuals, we compare the four simple models that rely on different tuning parameter values using data from the fourth collection. All of our models share nearly the same R-squared value and standardized estimates (see Table~\ref{tab:centr4_5-outD}). The negligible variance across these values provides no justification for choosing one parameter over another, meaning that giving more weight to repeated interactions with the same individuals makes no difference in our models. This suggests that the weighted network data is no more informative for anxiety shifts than the simple, binary network would be. The practical implications of this observation will the discussed in Sec.~\ref{sec:summary}. For that reason, we choose to test our final model using outdegree with $\alpha=0.0$, i.e., the standard version of degree that does not take frequency of repeated interactions into account. 

\subsubsection*{\bf\emph{Phase IV: Determining the final model}}

Our final linear regression model takes a variety of control variables into account, as per prior literature. Our control block includes anxiety at the beginning of the semester, i.e., pre-course scores ($pre.anxiety$), a binary gender variable (female or male, $gender$), and final course grade ($final.grade$): 

\begin{equation*}
\begin{split}
M_{full} : anxiety.shift \sim &centrality + gender \\
& + final.grade + pre.anxiety
\end{split}. 
\end{equation*}

We find that, regardless of students' anxiety at the beginning of the semester, gender, and final course grade, outdegree  with $\alpha=0.0$ is a significant and negative predictor of physics anxiety shift (standardized estimate $\beta=-0.19$, standard error of the standardized estimate $SE \beta=0.08$, $t$-test statistics $t=-2.47$, significance level $p<0.05$). Gender is also a significant predictor of students' shift in physics anxiety and male students are more likely than female students to experience a decrease in anxiety ($\beta=-0.25$, $SE \beta=0.08$, $t=-3.25$, $p<0.01$). As expected, the most significant effect on the anxiety shift comes from the pre-course anxiety score ($\beta=-0.41$, $SE \beta=0.08$, $t=-5.42$, $p<0.001$) and the final grade ($\beta=-0.35$, $SE=0.08$, $t=-4.54$, $p<0.001$). However, to have information about final grades one has to wait until the end of the semester, at which point no intervention is possible. Thus, we test our model with the $final.grade$ factor removed. As can be seen in Table~\ref{tab:centr4-simplified}, in the absence of final grades data, the outdegree measure and pre-course anxiety become the most significant predictors for anxiety shift. For every one standard deviation increase in a student's outdegree, their normalized physics anxiety would decrease by $0.29$ standard deviation.  

\begin{table}[t]
\renewcommand{\arraystretch}{1.1}
\renewcommand{\tabcolsep}{8pt}
\caption{Summary of the simplified linear regression model for anxiety shift with outdegree centrality from fourth collection ($\alpha=0.0$) and with the $final.grade$ factor removed: the standardized estimate ($\beta$), the standard error for the standardized estimate (SE $\beta$), and  $t$ test statistic ($t$). We consider networks without instructional staff. Significant $p$-values are marked with an asterisk.}
\begin{tabular}{c|ccc} \hline\hline
Factor & $\beta$ & SE $\beta$ & $t$ \\ \hline
$C_{outD}^{0.0}$ & $-0.29^{***}$ & $0.08$ & $-3.56$ \\ 
Gender (M)       & $-0.24^{**}$  & $0.08$ & $-2.92$ \\ 
Pre-anxiety      & $-0.37^{***}$ & $0.08$ & $-4.53$ \\
\hline \hline
\multicolumn{4}{l}{\footnotesize ***$p<0.001$, **$p<0.01$}
\end{tabular}
\label{tab:centr4-simplified}
\end{table}

\section{Discussion}\label{sec:discussion}
We start our exploration of the relationship between students' classroom interactions and their anxiety by looking at changes the latter. Students' average {\it pre} and {\it post} physics anxiety scores exhibit no statistical differences, yet the data and its distribution indicate that while overall shift does not occur, individual shifts do. Some students experience increases in anxiety, while others experience decreases. We want to better understand the factors that might contribute to these changes. Prior research in active-learning physics classrooms indicate that student self-efficacy, a construct related to anxiety, correlates with the kinds of classroom interactions students participate in~\cite{Dou16-BPM}. Moreover, the broader literature on anxiety suggests that student behavior and classroom participation has reciprocal relationships with anxiety~\cite{Paivio59-MCA,Hills07-ICR,Camacho95-AGB}. 

We quantify the social integration of students in the classroom using the tools of SNA. After surveying students regarding the meaningful academic interactions they participated in, the list of interactions derived from their responses are used to calculate three important measures of individuals' relational position in the networks: indegree, outdegree, and closeness. Simple linear models between students' normalized shifts in physics anxiety and each of these centrality measures reveals a significant relationship only for the outdegree: the more interactions students report having, the more likely they are to experience a decrease in physics anxiety. Given the correlational nature of these models, we would also expect students whose anxiety decreases over time to report a greater number of meaningful academic interactions. 

The relationship between physics anxiety and classroom interactions is meaningful, given the overall trend towards active learning modalities in physics teaching. Research suggests that for some students, active learning environments may cause discomfort and anxiety~\cite{Villasenor07-RPI,Turpen10-CDC,Dou18-USE}, which can lead to suppressed performance or loss of interest -- factors that affect persistence in a major~\cite{Lent94-UST}. Physics instructors that solicit peer learning must take into consideration a variety of ways to group students in order to optimize outcomes like learning and improved attitudes towards physics. Given the relationship between these factors and anxiety, our study suggests students should be given opportunities to interact with as great a number of peers as possible. 

Students' outdegree can be interpreted in two ways. It can be thought of as the number of interactions the student in question actively engages in. This interpretation assumes that the student is exercising agency in their interactions, listing peers they purposefully sought after. Overall trends in network data from this and similar physics classrooms suggest this to be the case~\cite{Bruun14-TDN,Dou17-PHD}. The other possible interpretation does not necessarily imply a form of student agency, but rather considers student perception instead. Students who perceive having had more meaningful interactions, regardless of whether they initiated these interactions or not, list these interactions on a survey and, as a result, have greater outdegree centrality than those who do not perceive having as many meaningful interactions. This interpretation suggests a reciprocal relationship between anxiety and the number of meaningful interactions students perceive having. When taking this latter interpretive approach, interactions listed may include passive events where the student was the subject of someone's initiative rather than the actual initiator. We find this unlikely to be the case given that indegree, a truly passive measure of which the student has no control, was not a significant predictor of anxiety shifts. In other words, simply being the subject of others' interactions is not related to anxiety shifts. More likely, students must initiate the interaction in at least some of the cases in order to benefit from the relationship between outdegree centrality and negative shifts in physics anxiety. Regardless of one's interpretation, the act of identifying and listing meaningful interactions must be taken by the student.

Our analyses also indicate that when exploring student interactions in the physics classroom, the advantage provided by taking into account the frequency of repeated interactions between the same two individuals is relatively small. A comparison of beta estimates and R-squared values reveals only minor differences between the effect size of outdegree, regardless of whether we used a tuning parameter that did not take repeated interactions into account (i.e., $C_{outD}^{0.0}$) or one that greatly advantaged students with repeated interactions (i.e., $C_{outD}^{1.0}$; see Table~\ref{tab:centr4_5-outD}). No other study examining classroom interactions has compared the outcomes of not taking repeated interactions into account versus doing so. Given the extra cognitive effort required for students to recall the repeatedness of interactions, as well as the additional work involved in both collecting and analyzing this type of data, it seems that the frequency of interactions can be ignored (unless prior literature indicates a potential increased effect).  

On the other hand, students' self-reported gender, pre-course anxiety and final grade in the course all significantly contribute to predicting students' shifts in anxiety. As expected, male students are more likely to experience decreases in anxiety, as are students who finished the semester with higher grades~\cite{Williams01-UAG,Leppavirta11-PHD,Tobias85-TA}. Students with higher outdegree measured sometime after the second midterm are also more likely to experience decreases in physics anxiety. 

Of all these variables, outdegree lends itself most readily to direct intervention design given that it can be easily measured and, unlike final grades, plays a role long before the semester ends. Instructors can help students feel less anxious by creating an environment that fosters and invites social interactions related to the content. We should note that students in these classrooms, on average, reported interacting with more than just their group members. Average outdegree during the fourth and fifth collection is 4.74 ($SD=4.50$) and 5.73 ($SD=4.91$), respectively, despite the fact that students were organized in groups of three. The class structure welcomes and sometimes invites students to interact across groups, which has also been associated with increased learning~\cite{Rienties17-TGI}. Thinking carefully about how to invite and solicit positive academic interactions will help decrease students' physics anxiety regardless of their academic performance or incoming anxiety levels. Our social network approach suggests that fixing groups and/or forcing students to work only within established groups may not support a positive learning environment.

\section{Summary}\label{sec:summary}
SNA not only provides a novel set of tools that can help physics education researchers better understand how social interactions contribute to other factors, it can also be used in practical ways to assess social dynamics. In this study a simple count of who interacted with whom would not have drawn out the nuance provided by differentiating outdegree from indegree. Moreover we would not have concluded that closeness, the most significant and meaningful centrality measure in terms of predicting students' persistence~\cite{Zwolak18-PIN}, is not related to changes in physics anxiety. Our use of SNA makes sense given our research questions, and our outcomes lead to practical recommendations for active-learning physics classrooms. In the case of physics anxiety, instructors can use simple SNA surveys throughout the semester to gauge what kind of interactions their classroom structure is fostering. This data can be used to quickly calculate student centrality using programs like R that automate the majority of the process. Interventions can then be designed to encourage the kinds of interactions that maximize positive learning experiences.

Finally, we encourage researchers to think broadly about the potential uses of SNA in research. While we focus here on the classroom environment, SNA can be applied to studies of informal learning environments, as well. These kinds of settings do not necessarily take place in a physical space either. Mobile phone applications like Whatsapp and Messenger are often used by students outside of class to share information and organize meetings. These virtual communication tools lend themselves to exploration via SNA. Moreover, social networks do not necessarily have to involve direct interactions, but can be defined to capture physical proximity networks, attendance-absence networks, or networks defined by non-verbal cues, to name a few. We believe that the growing prominence of active-learning strategies and the relationship between social interactions and student success will further require the use of SNA to help improve student persistence and retention. Implementing the suggestions here gives the ultimate test of their efficacy.

\appendix
\section{The normalized gain}\label{app:nor_gain}
Since its introduction in 1998, the normalized gain has been commonly used as a measure of students averaged improvement over time in various context. Defined as a measure of the ``average effectiveness of a course in promoting conceptual understanding''~\cite{Hake98-MIP}, it is typically used to capture the average trends for the entire class. By adjusting values measured on different scales, it also allows comparison between different groups. However, the normalized gain is not robust when a large drop in scores takes place. 

For simplicity, lets assume that the scores range from 0 to 100\,\%. The normalized gain on an individual level is defined as:
\begin{equation}
g_{norm} = \frac{post - pre}{100 - pre},
\end{equation}
where {\it pre} and {\it post} denote the pre- and post-course scores, respectively. For averaged gain, as introduced in Ref.~\cite{Hake98-MIP}, {\it pre} and {\it post} need to be replaced by the respective averages over the entire class, i.e., $\langle pre\rangle$ and $\langle post\rangle$  While this equation always yields values smaller or equal to one (simply because {\it post} can be at most 100), when {\it post} score is lower than {\it pre} score (i.e., when a drop in scores rather than gain is observed), it is possible to see values $g_{norm}<-1$. This happens if
\begin{equation*}
post < 2(pre-50),
\end{equation*}
that is if, after scoring more than $50\,\%$ on the pre-test, an individual has a {\it post} score of no more than $2(pre-50)$.

While such big differences are less likely when {\it pre} and {\it post} scores are averaged over the entire class, it is still possible to see a ``normalized gain'' that is outside of $[-1,1]$ range, invalidating the comparison between sections. However, this lack of robustness against large drops in scores should not be thought of as an argument against using the normalized gain. On the contrary, this property of $g_{norm}$ provides researchers with a tool for quick detection of atypical performances and possible outliers (e.g., students who did not give genuine responses on the post-course data collection). We do argue, however, that a distribution of individual gains should be considered in addition to comparing the normalized gain values. As can bee seen in our data, majority of students did experience a shift in their anxiety, either positive or negative. However, had we railed solely of the measure of normalized shift, we would find no differences as the traditional normalized shift for our data is less than $0.3\,\%$ (see Fig.~\ref{fig:anx-dist} for the distribution of normalized shifts at individual level). This is particularly important when normalized gain is used to assess the effectiveness of a novel learning approach in smaller classroom, where few outliers can significantly affect the normalized gain.

\section{Descriptive statistics for centralities}\label{app:desc_centr}
\begin{table}[h!]
\renewcommand{\arraystretch}{1.1}
\renewcommand{\tabcolsep}{9pt}
\caption{The summary of the descriptive statistics for the outdegree centrality from fourth collection ($N=53$). Based on Shapiro-Wilk test, the null hypothesis about the normal distribution is rejected for all centralities. The median and interquartile range (IQR) are used to describe the distribution and dispersion for each measure. Note that instructional staff is removed from the network.}
\centering
\begin{tabular}{l|cccc} \hline \hline
\multirow{2}{*}{Centrality} & \multicolumn{2}{c}{Shapiro-Wilk test} & \multirow{2}{*}{Median} & \multirow{2}{*}{Mean} \\
  & $W$ & $p$  &  &   \\ \hline
$C_{outD}^{0.0}$  & 0.890 & $<0.001$ &  5.0 &  7.0 \\
$C_{outD}^{0.5}$  & 0.896 & $<0.001$ &  7.2 & 10.2 \\ 
$C_{outD}^{1.0}$  & 0.897 & $<0.001$ & 11.0 & 15.0 \\
$C_{outD}^{1.25}$ & 0.896 & $<0.001$ & 13.9 & 19.7 \\
\hline \hline
\end{tabular}
\label{tab:stats_centr_4}
\end{table} 

\begin{table}[h!]
\renewcommand{\arraystretch}{1.1}
\renewcommand{\tabcolsep}{9pt}
\caption{The summary of the descriptive statistics for all centrality measures from fifth collection ($N=53$). Based on Shapiro-Wilk test, the null hypothesis about the normal distribution is rejected for outdegree and closeness, but not for indegree. The median and interquartile range (IQR) are used to describe the distribution and dispersion for each measure. Note that instructional staff is removed from the network.}
\centering
\begin{tabular}{l|cccc} \hline \hline
\multirow{2}{*}{Centrality} & \multicolumn{2}{c}{Shapiro-Wilk test} & \multirow{2}{*}{Median} & \multirow{2}{*}{IQR} \\
  & $W$ & $p$  &  &   \\ \hline
$C_{inD}^{0.0}$  & 0.980 & 0.056 & 6.0 & 3.0 \\
$C_{inD}^{0.5}$  & 0.994 & 0.843 & 8.8 & 4.9 \\
$C_{inD}^{1.0}$  & 0.991 & 0.613 & 13.0 & 8.0\\
$C_{inD}^{1.25}$ & 0.991 & 0.574 & 15.2 & 10.0\\
\hline
$C_{outD}^{0.0}$  & 0.901 & $<0.001$ & 5.0 & 5.8 \\
$C_{outD}^{0.5}$  & 0.906 & $<0.001$ & 8.2 & 8.8 \\
$C_{outD}^{1.0}$  & 0.908 & $<0.001$ & 12.0 & 13.8 \\
$C_{outD}^{1.25}$ & 0.909 & $<0.001$ & 15.2 & 17.5 \\
\hline
$C_{C}^{0.0}$  & 0.789 & $<0.001$ &  0.42 & 0.08 \\
$C_{C}^{0.5}$  & 0.827 & $<0.001$ &  0.43 & 0.09 \\
$C_{C}^{1.0}$  & 0.866 & $<0.001$ &  0.43 & 0.11 \\
$C_{C}^{1.25}$ & 0.877 & $<0.001$ &  0.44 & 0.11 \\
\hline \hline
\end{tabular}
\label{tab:stats_centr_5}
\end{table} 

R.D. and J.P.Z. contributed equally to this work.



\end{document}